\pdfoutput=1

\documentclass[11pt,twoside,a4paper,cmspaper,final,collab]{cms-tdr}
\def\svnVersion{66380:66601}\def\cmsCernNo{2011-081}\def\cmsCernDate{\today}\def\cmsMessage{Submitted to Physics Letters B}

\begin{document}\cmsNoteHeader{EXO-10-016}

\hyphenation{had-ron-i-za-tion}
\hyphenation{cal-or-i-me-ter}
\hyphenation{de-vices}
\RCS$Revision: 66601 $
\RCS$HeadURL: svn+ssh://alverson@svn.cern.ch/reps/tdr2/papers/EXO-10-016/trunk/EXO-10-016.tex $
\RCS$Id: EXO-10-016.tex 66601 2011-07-09 11:36:03Z alverson $
\newcommand{\pho}{\phantom{0}}
\newcommand{\phoo}{\phantom{00}}
\cmsNoteHeader{EXO-10-016} 
\title{\texorpdfstring{A search for excited leptons in ${\rm pp}$ collisions at $\sqrt{s}= 7~\TeV$}{A search for excited leptons in pp collisions at sqrt(s)= 7 TeV}}

\address[neu]{Northeastern University}
\address[fnal]{Fermilab}
\address[cern]{CERN}
\author[cern]{The CMS Collaboration}

\date{\today}

\abstract{
A search for excited leptons is carried out with the CMS detector at the LHC, using
$36~\mathrm{pb}^{-1}$ of
{\mbox{${\rm pp}$}}\ collision data recorded at $\sqrt{s}= 7 \TeV$.
The search is performed for  associated
 production of a lepton and an oppositely charged excited lepton
 $ \mathrm{pp} \rightarrow  \ell \ell^*$, followed by
the decay $\ell^* \rightarrow  \ell \gamma$, resulting in the
$\ell \ell \gamma$ final state, where $\ell = \mathrm{e},\mu$.
No excess of events above the standard model
 expectation is observed.
Interpreting the findings in the context of $\ell^*$~production through four-fermion contact
interactions
and subsequent decay via electroweak processes, first
 upper limits are reported
for $\ell^*$ production at this collision energy. The exclusion region
 in the compositeness scale $\Lambda$ and excited lepton mass $M_{\ell^*}$
parameter space is extended beyond  previously established limits.
For $\Lambda = M_{\ell^*} $,
excited lepton masses are excluded below 1070~\GeVcc
for $\mathrm{e}^*$ and 1090~\GeVcc for ${\mu}^*$ at the $95\%$ confidence level.
}

\hypersetup{%
pdfauthor={CMS Collaboration},%
pdftitle={A search for excited leptons in pp Collisions at sqrt(s) = 7 TeV},%
pdfsubject={CMS},%
pdfkeywords={CMS, physics}}

\maketitle 

\section{Introduction}
\label{sec:introduction}

A fundamental question in the standard model (SM) of particle physics
concerns the source of the mass hierarchy of
quarks and leptons. 
 A possible explanation for the three generations is a compositeness model
in which the known leptons and quarks are  bound
states of either three fermions or  a
fermion-boson pair~\cite{compositeness1}.
This substructure, if it exists, implies
a large spectrum of excited states. 
Novel strong interactions would couple the excited fermions to ordinary quarks and leptons.
These contact interactions  can be described with an effective Lagrangian \cite{Baur90}:
\begin{equation*}
\label{BaurEq1}
 \mathcal{L}_{\mathrm{CI}}  = \frac{g^{*2}}{2\Lambda^2}\, j^{\mu}\, j_{\mu},
\end{equation*}
where $\Lambda$ is the compositeness or contact interaction scale,
 $g^{*2}$ represents a coupling constant chosen to be
 $4 \pi$, and $j_{\mu}$ is the fermion current:
\begin{equation*}
  j_{\mu}  =   \eta_L \, \bar{f}_L\gamma_{\mu}f_L + \eta'_L \,
  \bar{f}^*_L\gamma_{\mu}f^*_L + \eta''_L \, \bar{f}^*_L
  \gamma_{\mu}f_L
   + \, \text{h.c.} + \, (L \rightarrow R).
\end{equation*}
The SM and excited fermions are denoted by $f$ and $f^*$,
respectively. The subscripts $L$~($R$) refer to left- (right-) handed
fermions. The $\eta$ factors for left-handed currents are
conventionally set to one, and those for right-handed currents are set to zero.
Gauge-mediated transitions between ordinary and excited fermions are described by an
effective Lagrangian \cite{Baur90,boudjema93}:
\begin{equation}
 \label{BaurEq2}
  \mathcal{L}_{\mathrm{GM}} =  \frac{1}{2\Lambda}
  \bar{f}_R^*\,\sigma^{\mu\nu}
      \left[ g_s {\rm{f}}_s \frac{\lambda^a}{2} G^a_{\mu\nu} + g {\rm{f}} \frac{\tau}{2} W_{\mu\nu}
                    +   g'{\rm{f}}'\frac{Y}{2}B_{\mu\nu}\right] f_L + \text{h.c.,}  \nonumber
\end{equation}
where $G^a_{\mu\nu}$, $W_{\mu\nu}$, and $B_{\mu\nu}$ are the field strength tensors
of the gluon,  SU(2), and U(1) gauge fields, respectively,
and $g_s$, $g$, and $g'$ are the corresponding gauge couplings.
The scaling parameters ${\rm{f}}_s$, $\rm{f}$, and $\rm{f}'$ are assumed to be equal to one.
Previous searches at
LEP~\cite{Buskulic:1996tw, Abreu:1998jw, Achard:2003hd, Abbiendi:1999sa,delphi},
 HERA~\cite{H1estar,zeus},
and the Tevatron~\cite{cdfmu,d0,CDFestar,D0estar} have found no evidence
for such excited leptons.

This Letter presents a search for excited leptons in
 $\rm{pp}$ collisions at a centre-of-mass energy of 7~\TeV  collected in 2010
with the Compact Muon Solenoid (CMS)~\cite{JINST} detector at the
Large Hadron Collider (LHC). The data sample corresponds to an integrated luminosity of
36~${\rm pb}^{-1}$.
The production of an excited lepton $\ell^*$  ($\mu^*$ or $\mathrm{e}^*$) in association
with an oppositely charged lepton of the same flavour via four-fermion contact interactions,
followed by the electroweak decay $\ell^* \rightarrow \ell \gamma$, is considered. The resulting
final state, $\ell^+ \ell^- \gamma$, is fully reconstructed.
The dominant SM background for this search is the Drell--Yan production of $\ell^+\ell^-$ pairs,
 accompanied by a photon radiated either by an initial-state parton (ISR) or from one of the
final-state leptons (FSR).

\section{Experimental setup and event simulation}
\label{sec:samples}

The central feature of the CMS
apparatus is a superconducting solenoid of 6~m internal diameter,
providing a magnetic field of 3.8~T. Within the field volume are the silicon
pixel and strip trackers, the crystal electromagnetic calorimeter (ECAL), and
the brass/scintillator hadron calorimeter (HCAL).
The  central tracker
consists of 1440 silicon pixel and 15\,148 silicon strip detector modules.
It provides an impact parameter
resolution of approximately $15~$$\mu$m and a transverse momentum
($\pt$) resolution of 4\% for 500~\GeVc charged particles.
The ECAL has an energy resolution of better than 0.5\% above
100~GeV.
The calorimeter cells are grouped in projective towers of granularity
$\Delta \eta \times \Delta \phi = 0.087 \times 0.087$ in the region $|\eta|<$1.74
and $0.175 \times 0.175$ at higher values of $\eta$,
 where
the pseudorapidity $\eta$ is defined as $\eta = -\ln(\tan\frac{\theta}{2})$,
with $\theta$ being the polar angle with respect to the direction of the
counterclockwise beam, and
$\phi$  the azimuthal angle, both measured in radians.
A preshower
detector consisting of two planes of silicon sensors interleaved with
a total of 3~radiation lengths of lead  is located  in front of the ECAL, covering
$1.65 <|\eta|< 2.6$.
Muons are measured in
gas-ionization detectors embedded in the steel return yoke,
with detection planes made of three technologies: drift tubes in the barrel
region  ($|\eta|<1.2$), cathode strip chambers in the endcaps ($0.9 <|\eta|< 2.4$), and resistive plate
chambers covering both the barrel and the endcap
regions.  Matching the muons to the tracks
measured in the silicon tracker results in a transverse momentum
resolution between 1 and 5\% for $\pt$ values up to 1~\TeVc.
The CMS detector and its performance are described in detail in Ref.~\cite{JINST}.

Monte Carlo (MC) samples of the signal  process  are
produced with the {\sc Pythia}~\cite{Sjostrand:2006za} event generator, 
using the compositeness model described in~\cite{Baur90, boudjema93},
for different values of the $\ell^*$ mass. The signal samples are
normalized using leading-order cross sections.
The decay of excited leptons via contact interactions is taken
into account for the signal expectation by using the prescription
from Ref.~\cite{Baur90}.
As a cross-check, samples of the signal  process  are  also simulated
using a customized
version of the  {\sc CompHEP}~\cite{Boos:2004kh} event generator. For the event selection criteria used
in this analysis, event rates predicted by {\sc Pythia} and {\sc CompHEP} are found to agree to within
 about~2\%. This difference is taken as a systematic uncertainty on the predicted signal rate.
The dominant background process, the Drell--Yan production of $\ell^+\ell^-$ pairs with an ISR or FSR photon, is
simulated with the {\sc MadGraph}~\cite{Madgraph} event generator.
The expectation for this process is corrected with the prediction of the next-to-leading order
{\sc Baur} generator~\cite{Baur:1992cd}.
The {\sc Pythia} event generator is used to generate
samples for other SM background processes,
including  WW, WZ, ZZ,
$\mathrm{t\bar{t}}$, and, for the electron channel, $\gamma\gamma$ production.
All samples are generated  using the
CTEQ6L1~\cite{cteq} parametrization for the parton distribution
functions (PDF) and  passed through a detailed simulation of the CMS detector response
implemented with the GEANT4 package~\cite{GEANT4}.

\section{Event selection}

This section describes the  criteria used to select the events in the analysis.
Trigger and particle identification efficiencies and their statistical
uncertainties, determined via a tag-and-probe method~\cite{wzxs} using samples
of $\mathrm{Z} \rightarrow \ell^+\ell^-$ events, will be discussed below.

Events  are collected with  single-muon
and double-photon triggers.
Double-photon triggers require two electromagnetic clusters above a
$p_\mathrm{T}$ threshold, and  thus can be satisfied both by photons and electrons.
The trigger efficiency is about 99\% for $\mu^+\mu^-\gamma$ and close to 100\%  for
$\mathrm{e}^+\mathrm{e}^-\gamma$
events passing our final selection criteria.
The analysis accepts
events with one isolated photon, two isolated leptons with high
$p_\mathrm{T}$, and at least one reconstructed primary vertex.
In events containing more than one photon or
more than two leptons,
the highest-$p_\mathrm{T}$ objects are chosen.
Events containing particles from LHC machine-induced backgrounds,
such as beam halo and beam gas,
are rejected by requiring that the fraction
of high quality
tracks be at least 25\%
in events with
more than 10 tracks \cite{bib:Track}.

Details of the muon reconstruction and identification
are given elsewhere~\cite{bib:MuonsCRAFT,MuRec}.
When a track is found in the muon chambers (standalone muon), a matching
track in the central detector
is required. A fit combining hits from these two matching tracks is
performed, resulting in a "global-muon track".
Alternatively, a track in the central detector, loosely matching with
hits in the muon detectors after extrapolation
to the muon chambers, results in a "tracker muon".
The tracker muons, together with the corresponding global-muon tracks, are selected for
the analysis.
Cosmic rays, muons from decay in flight of hadrons, and hadrons
misidentified as muons are rejected using requirements on the quality of the global-muon fit,
number of detector layers with hits in the muon chambers, and transverse impact parameter of the
track.
Two isolated muons  with $p_\mathrm{T} > 20~\GeVc$ and $|\eta|< 2.4$ are used in the analysis.
 The reconstruction and identification efficiency for muons with
$p_\mathrm{T} > 20~\GeVc$ is $(96 \pm 1)\%$.

Electrons and photons are detected in the ECAL as localized clusters~\cite{PhoRec,EleRec}.
The electron and photon identification procedures exploit the
ECAL shower shape and  isolation variables, the relative energy fraction
deposited in the hadronic and electromagnetic calorimeters ($H/E$), and, for electrons, the
presence of a track matching the ECAL cluster.
Applying different selection criteria  to these variables separates
clusters originating from electrons, photons, and hadrons.

For electrons, a central-detector
track matching the ECAL cluster is required. The track parameters
are extrapolated to ECAL; the energy   and extrapolated positions are required
to be consistent with those of the ECAL cluster. Electrons are
identified both in the ECAL barrel ($|\eta|< 1.44$)  and  endcaps ($1.57 < |\eta| < 2.5$),
which are the regions covered by both the tracker and ECAL.
Events in the
 $\mathrm{e}^+\mathrm{e}^-\gamma$ channel are required to contain
at least two electrons with $p_\mathrm{T} > 25~\GeVc$.
This threshold excludes $p_\mathrm{T}$ regions with lower trigger efficiency. Electrons are selected
with an average efficiency of $(91.4 \pm 0.3)\%$ in the barrel and
$(90.6 \pm 0.6)\%$ in the endcaps.

Photons are selected as clusters in  the ECAL barrel.
To accept converted photons, no additional requirement based on the presence of
matching tracks is applied.
Only ECAL  clusters that have not been previously matched to either of the
two highest-$p_\mathrm{T}$ electrons can be identified as photons.
Isolated photons with $p_{\rm T} > 20 \GeVc$ are used.

The photon should be separated from each of
the selected leptons in the $\eta - \phi$
plane by $\Delta R>0.5$, where $\Delta R = \sqrt{ (\Delta \phi)^2 +(\Delta \eta)^2}$ and
 $\Delta \phi$ and $\Delta \eta$ are the azimuthal angle and pseudorapidity differences
between the photon and the lepton.
Only events with dilepton invariant mass $M_{\ell\ell} > 60~\GeVcc$
are selected for further analysis.

Of the two possible  lepton-photon invariant mass combinations in each event,
the higher value, $M^\mathrm{max}_{\mu\gamma}$ or
$M^\mathrm{max}_{\mathrm{e}\gamma}$, is used as the search variable.
The use of the second mass combination does not improve the search
 sensitivity for the range of excited lepton masses probed.

\section{Background estimation}

Irreducible backgrounds from SM processes and instrumental
backgrounds from events in which jets are mis-reconstructed as leptons or photons
are evaluated separately.
The contribution from
SM processes with real leptons and photons, dominated by   Drell--Yan with ISR/FSR production,
is estimated from MC simulation.
The predicted background yields are corrected to
account for the difference in the efficiencies measured from
data and simulated events. The scale
factors are $0.967 \pm 0.025$ for photons, $0.989 \pm 0.010$ for muons,
$0.978 \pm 0.004$ for electrons in the barrel, and $0.994 \pm 0.006$ for
electrons in the endcaps.

\begin{table*}[htbp]
\begin{center}
\caption{Predicted and observed numbers of events  passing all the selection
  criteria for the muon and electron final states. Columns $2-4$ list separately the
  numbers of background
  events from the following sources: final states containing two leptons
  and one photon, including the dominant process, Drell--Yan
  with ISR or FSR; final states with two leptons
  accompanied by a  jet misidentified as a photon; and final states with one genuine lepton and one
  genuine photon, accompanied by a
  jet misidentified as a lepton. Contributions from final states with two or three
  misidentified jets are found to be negligible. Statistical and systematic uncertainties
 are summed in quadrature.
\label{tab:final_background}}
\begin{tabular}{|c|c|c|c|c|c|}
\hline
Final State &  $\ell^+\ell^- \gamma $ &  $\ell^+\ell^-  + \mathrm{jet} $ &
$\ell \gamma  + \mathrm{jet} $   & Total & Observed \\
\hline
$\mu^+\mu^-\gamma$ &  $19.1 \pm 1.4$ & $5.5 \pm 2.1$ &  $0.7  \pm 0.9 $ &  $26.7 \pm 2.7$ & 25 \\
$\mathrm{e}^+\mathrm{e}^-\gamma$ &  $11.0 \pm 1.0$ &  $1.4 \pm 0.8$ &  $1.0 \pm 0.4 $ &  $13.4 \pm 1.4$ & 7 \\
\hline
\end{tabular}
\end{center}
\end{table*}

Backgrounds from processes in which jets are
misreconstructed as leptons or photons
are measured with data samples selected to contain predominantly
jets~\cite{LED,EleRec}.
For each jet-enriched data sample, the misidentification  rate is measured as the ratio of the number
of objects passing all selection cuts (numerator)
to the number of potentially misidentifiable objects (denominator).
For the muon misidentification rate estimation,  the denominator
corresponds to the number of tracker muons with $|\eta|<$ 2.4.
For electrons,  the denominator
is the number of ECAL clusters with $H/E<0.05$ and  $p_\mathrm{T}>20~\GeVc$.
For photons,  the denominator is the number of
photon candidates  obtained by relaxing the isolation or shower-shape criterion.

Signal search samples containing one or more potentially misidentifiable objects are selected and used
together with the measured misidentification rates in order to predict the number of background events
with misidentified objects.
A closure test of the misidentification-rate method was done
using MC simulation. Good agreement between the
expected and observed number of events is found.
The background prediction is also tested
by comparing the observed and expected numbers of events
in several data control regions: samples where only one
lepton and one photon are selected, and samples containing two leptons
and a photon  selected with a looser sets of criteria.

Table \ref{tab:final_background} compares the predicted and
observed numbers of events passing all selection requirements.
Because of the lower efficiency for particle identification and the stricter
$p_\mathrm{T}$ requirement, yields  in the electron channel
are lower than in the muon channel.

\begin{figure*}[htbp]
\includegraphics[angle=0,scale=0.36]{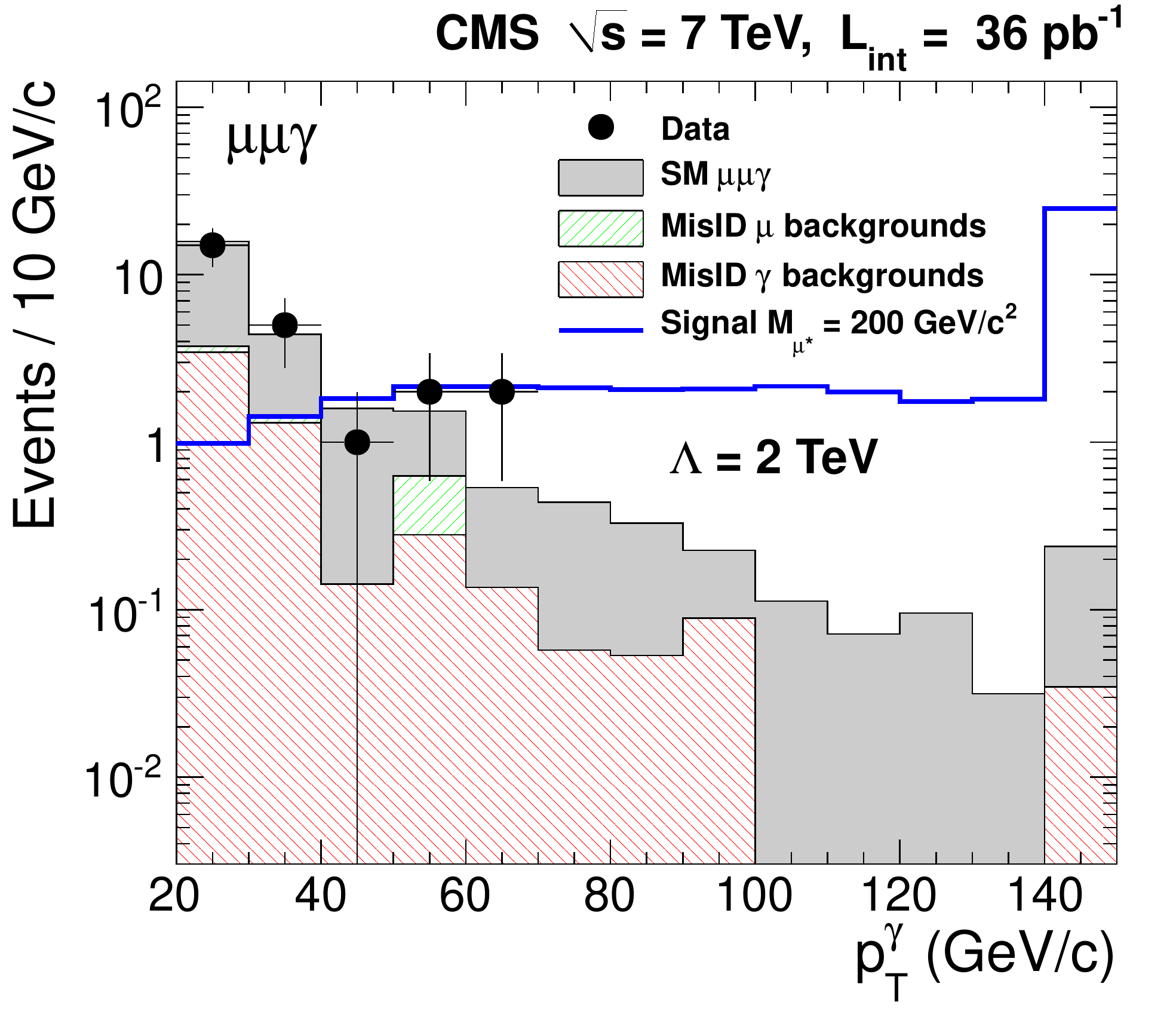} ~
 \includegraphics[angle=0,scale=0.36]{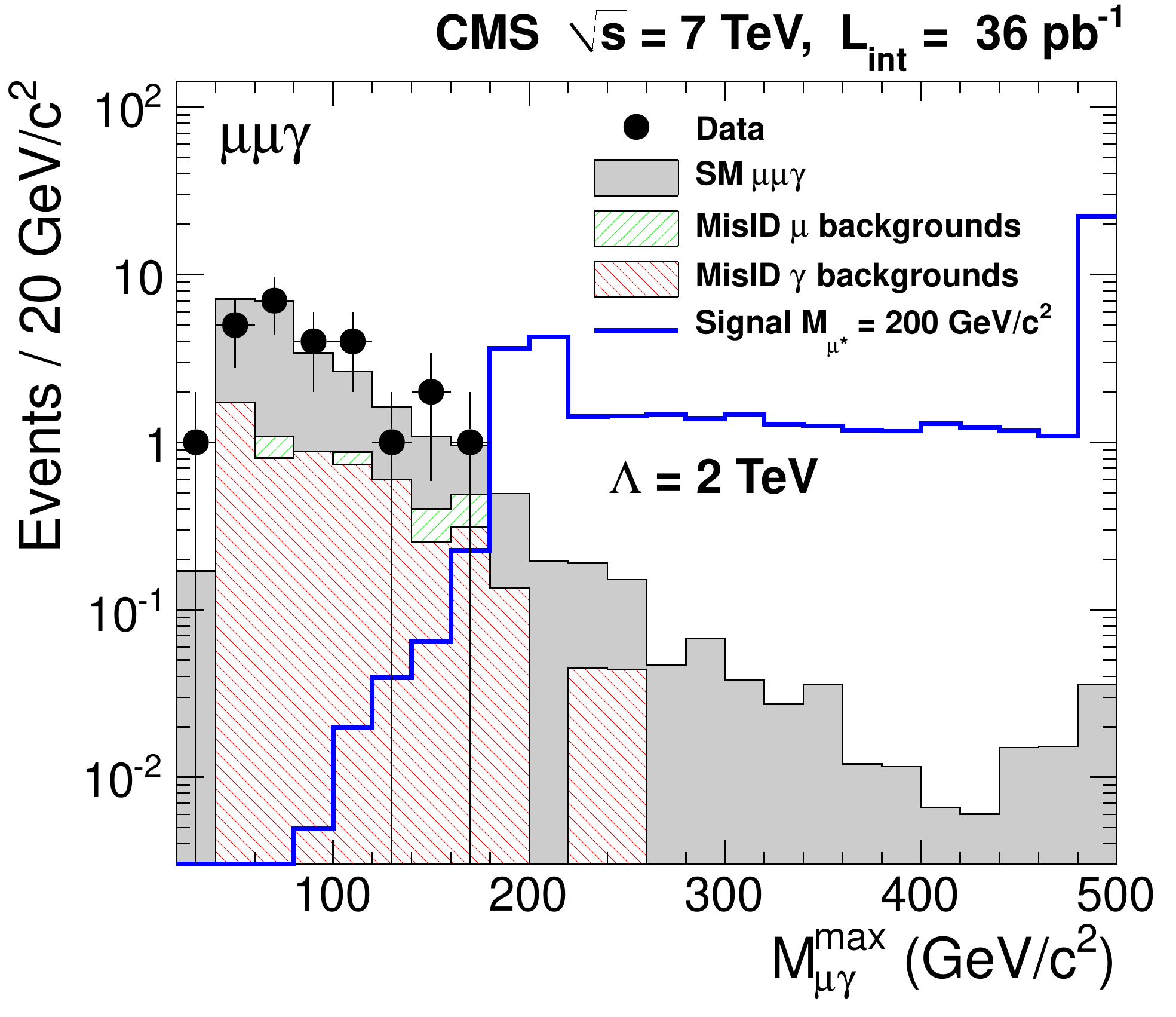}

\includegraphics[angle=0,scale=0.36]{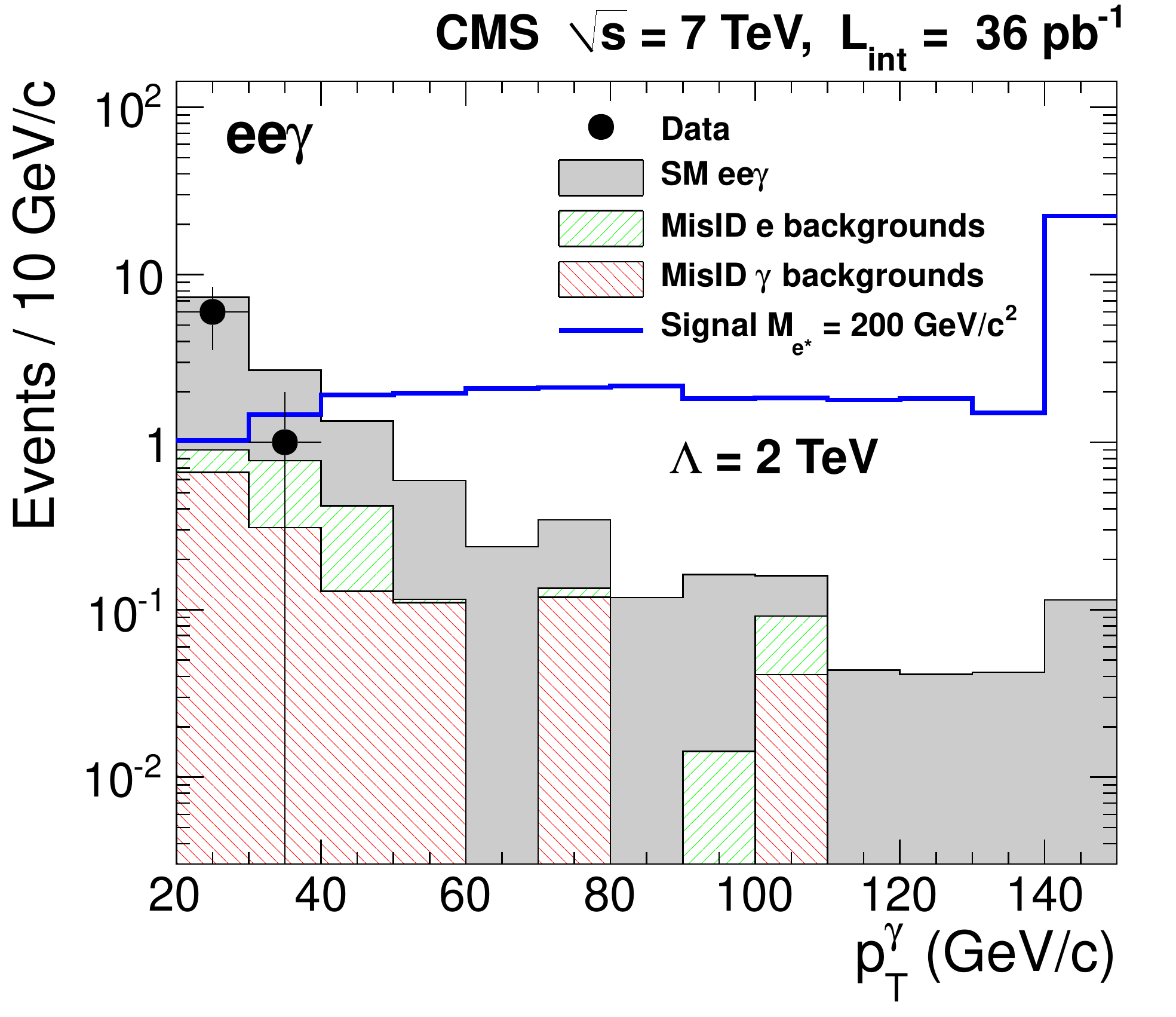} ~
\includegraphics[angle=0,scale=0.36]{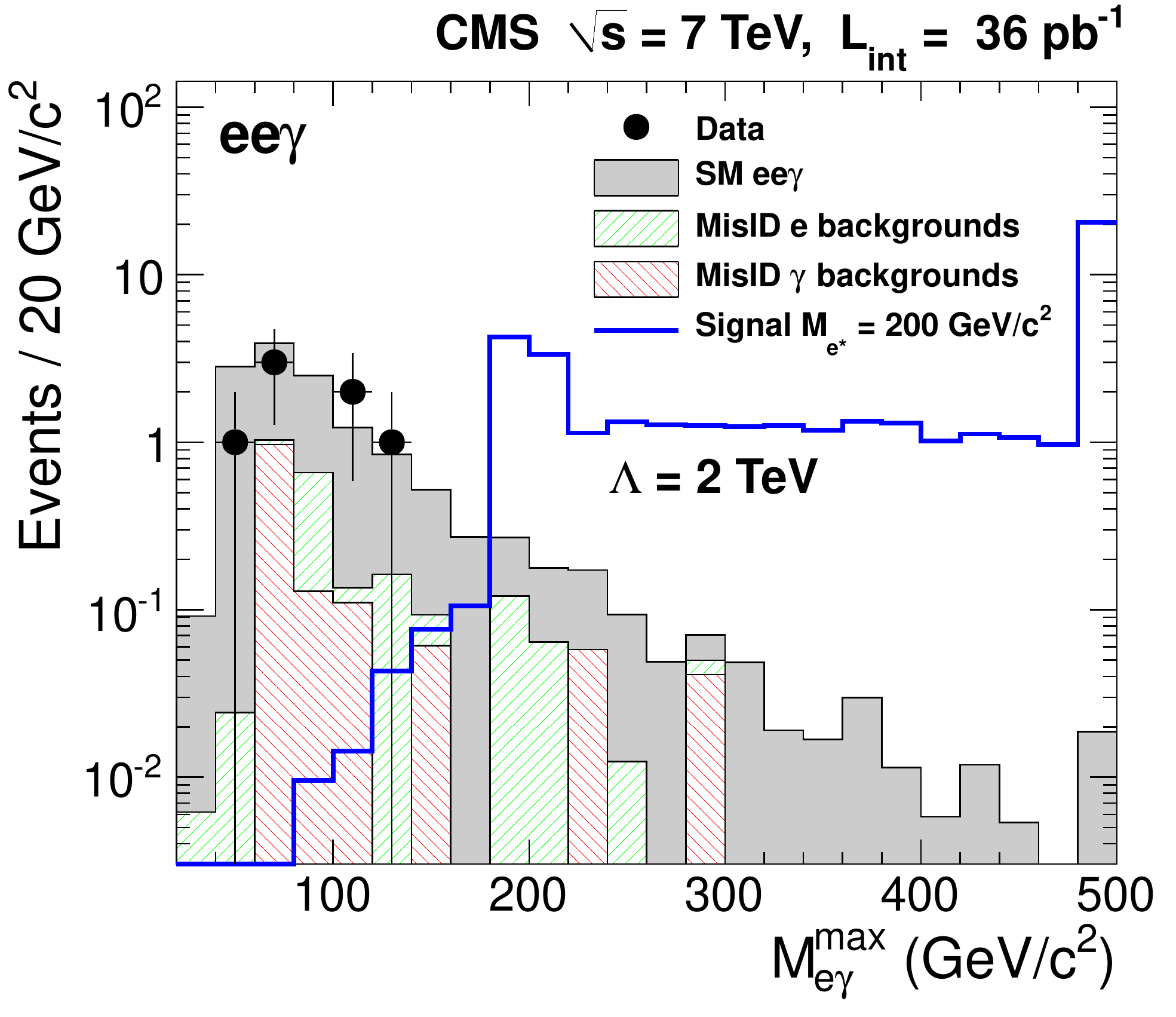}
\caption{
Photon transverse momentum  distributions (left) and maximum invariant mass distributions of the
  lepton-photon pair (right) in the muon (top) and
electron (bottom) channels. The data are shown as solid circles with error bars and the expected SM
background distributions are shown as hatched histograms.
The solid-line histogram displays the expected excited lepton signal
for $M_{\ell^*} = 200~\GeVcc$,
$\Lambda = 2 \TeV$. For this particular $M_{\ell^*}$,  the search region is restricted to
$M^\mathrm{max}_{\ell\gamma}>180~\GeVcc$. In each histogram, the
last bin includes the overflows.
}
\label{fig:mgMaxMass}
\end{figure*}

Figure~\ref{fig:mgMaxMass} (left) shows
the photon transverse momentum distributions in the electron (top) and muon (bottom)
channels, and (right)
the maximum lepton-photon invariant mass distributions
for data, along with the predictions for a signal with $M_{\ell^*} = 200$~\GeVcc,
$\Lambda = 2 \TeV$, and for the SM background
processes. The background prediction describes the data well in these
as well as in other kinematic variables.

\section{Systematic uncertainties}

The normalization and shape of the invariant mass distributions used to establish a
possible excited lepton signal are subject to uncertainties from both
experimental and theoretical sources.
The normalizations of
the spectra are based on
the integrated luminosity of the data sample, which is
known to a precision of 4\%~\cite{lumipas}.
The theoretical calculations of background process cross sections are affected
by uncertainties in parton distribution functions~\cite{cteq} and the choice of
factorization  and renormalization scales.
The uncertainties on the PDFs are evaluated using a reweighting technique with the CTEQ6M
parametrization~\cite{cteq}, while the uncertainties on the
factorization and renormalization scales are estimated by varying them simultaneously from
half to twice their central values. The resulting uncertainty on the background expectation is
found to be~5\%.
In this section, all quoted uncertainties  are
obtained after requiring $M^\mathrm{max}_{\ell\gamma}>180~\GeVcc$.

The uncertainty on the number of background events from jets misidentified as leptons or photons
 is estimated by comparing misidentification rates measured in jet-enriched samples collected with
 different trigger requirements. Another source of uncertainty, estimated using MC
 simulations, is the difference between the misidentification rate observed in the jet-triggered
 samples, where it is measured, and the photon- or muon-triggered samples, where it is applied.
The photon misidentification-rate uncertainty increases from 20\% to~50\% with photon $p_\mathrm{T}$.
This results in an uncertainty of 10\%~(7\%) on the background prediction
in the muon (electron) channel.
The electron misidentification rate is known with a 25\% (40\%) uncertainty in the ECAL barrel
 (endcaps), resulting in a 10\% uncertainty on the background prediction in the
 electron channel. The uncertainty on the muon misidentification rate is estimated to be~50\%,
 and the resulting effect on the background expectation is~1\%.

The uncertainties on the efficiency correction factors used in
simulated events are included in the 
systematic uncertainties. They   are 0.6\% (1.1\%) for electrons measured in the ECAL barrel (endcaps),
1\% for muons,   and 2.5\% for photons.
The effect on signal and background yields due to the particle identification
uncertainties is smaller than 2\% for the leptons and about 2.5\% for the photons.
The uncertainty on the photon and electron energy scale translates into an additional
 uncertainty of
0.5\% for signal and 1.2\% for background predictions.

Considering all sources of uncertainties mentioned above, the
selection efficiencies for excited leptons are known to a  precision of
$3-4\%$, in both  the electron and muon channels.

\section{Results and discussion}

In order to enhance the sensitivity of the analysis, the search is restricted to
the high invariant mass region by applying a selection on
$M_{\ell\gamma}^{\rm max}$  that depends
on the excited lepton mass hypothesis. For each excited lepton mass,
the entire analysis is repeated using various search regions and the
region giving the best expected limits is taken.
For excited lepton masses above
600~\GeVcc, where almost no background is expected,  the search region
$M_{\ell\gamma}^{\rm max}>500$~\GeVcc is used.
The number of observed events, and the predicted numbers from SM backgrounds
and from an excited lepton signal, as well as the
selection efficiency, are listed in
Table~\ref{tab:muresults} for the two search channels with
various excited lepton mass hypotheses and fixed $\Lambda = 2$~TeV.  The uncertainties
on the number of predicted background events and the
signal efficiencies are
the statistical and systematical uncertainties summed in quadrature.
No excited lepton candidate events are found in any of the search regions.
This lack of events is consistent with the SM background predictions.

\begin{table*}[htbp]
   \begin{center}
     \caption{
Minimum requirement on the highest invariant mass pair
$M^\mathrm{cut}_{\ell\gamma}$,
  number of events observed in data and expected SM background,
  signal efficiency, observed (expected)
upper limits $\sigma^\mathrm{lim}_\mathrm{obs}$
($\sigma^\mathrm{lim}_\mathrm{exp}$)
on the $\ell^*$ production cross section times the branching
fraction of the $\ell^* \rightarrow \ell \gamma$ decay, and
expected numbers of signal events,
for various excited lepton masses $M_{\ell^*}$,
assuming $\Lambda = 2\TeV$.
 Invariant masses are given in~\GeVcc and  $\sigma^\mathrm{lim}_\mathrm{obs}$
($\sigma^\mathrm{lim}_\mathrm{exp}$) in~pb. The uncertainties
displayed correspond to statistical and systematical uncertainties summed in quadrature.}
  \label{tab:muresults}
     \begin{tabular}{|c|c|c|c|c|c|c|c|}     \hline
\multicolumn{7}{|c|}{Muon Channel} \\ \hline
       $M_{\mu^*}  $ & $M^\mathrm{cut}_{\mu\gamma}$  & N$_{\rm data}$ &
       N$_{\rm predicted\ SM}$  & Signal eff.\ (\%)  &
       $\sigma^\mathrm{lim}_\mathrm{obs}\ (\sigma^\mathrm{lim}_\mathrm{exp}) $ & N$_{\rm predicted\
         signal}$    \\\hline
\pho    200 & 180 & 0 & 1.35 $\pm$ 0.20 & 44.8 $\pm$ 1.8 & 0.19 (0.28)  & 47 \\
\pho    400 & 350 & 0 & 0.11 $\pm$ 0.09 & 51.0 $\pm$ 1.9 & 0.16 (0.17)  & 18.6 \\
\pho    600 & 500 & 0 & 0.04 $\pm$ 0.08 & 53.9 $\pm$ 2.0 & 0.15 (0.15)  & 7.3 \\
\pho    800 &  500 & 0  & 0.04 $\pm$ 0.08 & 55.6 $\pm$ 2.1 & 0.15 (0.15)  & 2.8 \\
        1000 &  500 & 0 & 0.04 $\pm$ 0.08 & 56.9 $\pm$ 2.1 & 0.15 (0.15)        & 1.1 \\
        1200 &  500 & 0 & 0.04 $\pm$ 0.08 & 56.9 $\pm$ 2.1 & 0.15 (0.15)        & 0.4 \\
        1500 &  500 & 0 & 0.04 $\pm$ 0.08 & 58.5 $\pm$ 2.1 & 0.14 (0.14)    & 0.1 \\
\hline
\multicolumn{7}{|c|}{Electron Channel} \\
\hline
       $M_{\mathrm{e}^*}  $ & $M^\mathrm{cut}_{\mathrm{e}\gamma}$  & N$_{\rm data}$ &
N$_{\rm predicted\ SM}$
& Signal eff.\ (\%)  &   $\sigma^\mathrm{lim}_\mathrm{obs}\ (\sigma^\mathrm{lim}_\mathrm{exp}) $ &
N$_{\rm expected\ signal}$    \\\hline
\pho 200  &  180 &  0 &  1.00  $\pm$ 0.15 & 38.7  $\pm$ 1.5 &  0.21 (0.30)  &40 \\
\pho 400  &  350 &  0 & 0.10   $\pm$ 0.07 & 44.6  $\pm$ 1.7 &  0.19 (0.19) &16 \\
\pho 600  &  500 &  0 & 0.01   $\pm$ 0.06 & 47.0  $\pm$ 1.7 &  0.18 (0.18)  &6.4 \\
\pho 800  &  500 &  0 & 0.01   $\pm$ 0.06 & 49.3  $\pm$ 1.8 &  0.17 (0.17)  &2.5 \\
     1000  &  500 &  0 &  0.01   $\pm$ 0.06 & 50.9  $\pm$ 1.8 &  0.16 (0.16)       &1.0 \\
     1200  &  500 &  0 &  0.01   $\pm$ 0.06 & 51.3  $\pm$ 1.8 &  0.16 (0.16)       &0.4 \\
     1500  &  500 &  0 &  0.01   $\pm$ 0.06 & 52.9  $\pm$ 1.8 &  0.16 (0.16)       &0.1 \\
\hline
\end{tabular}
\end{center}
\end{table*}

\begin{figure*}[htbp]
  \begin{center}
    \includegraphics[angle=0,width=0.49\linewidth]{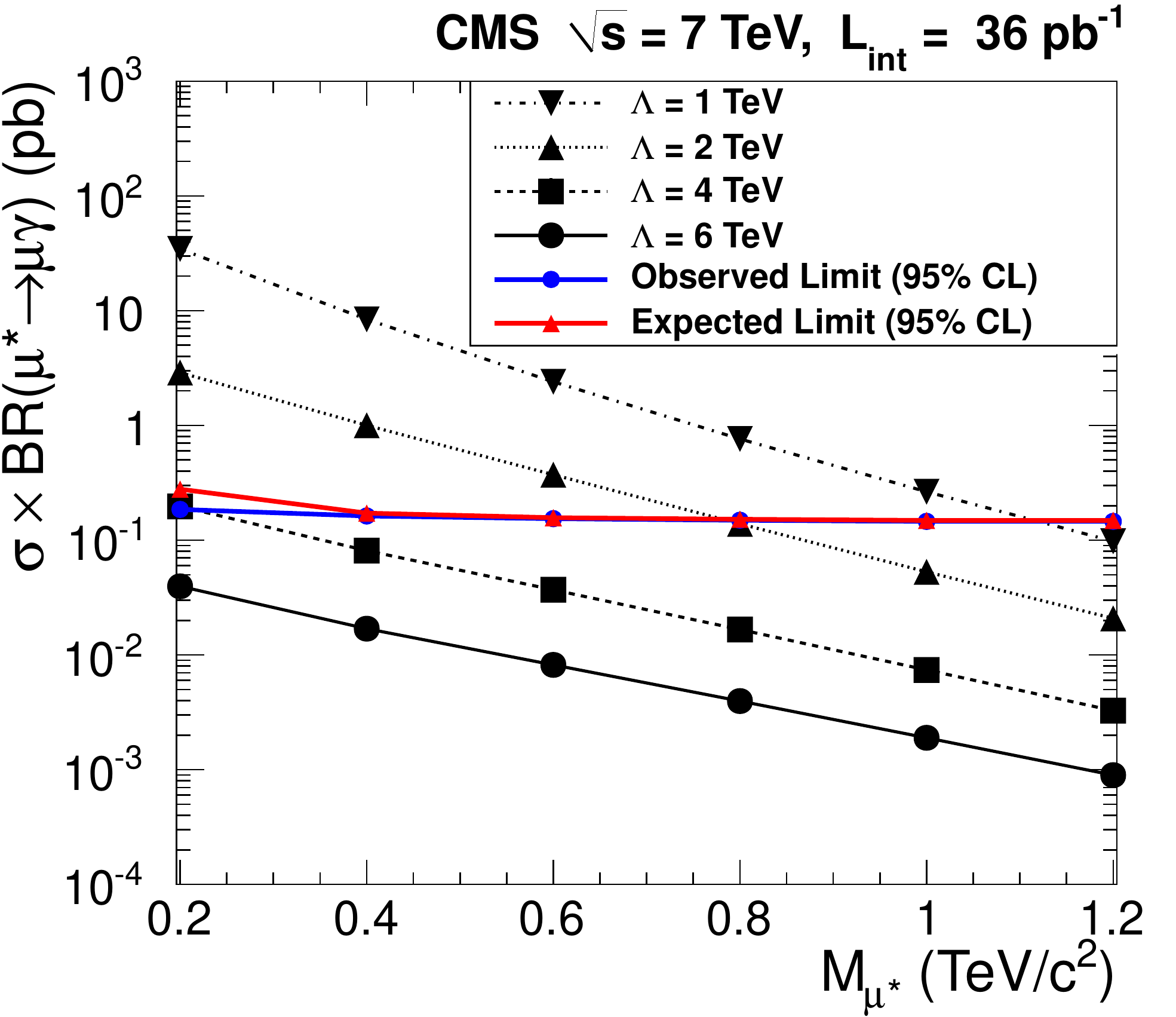}
    \includegraphics[width=0.49\linewidth]{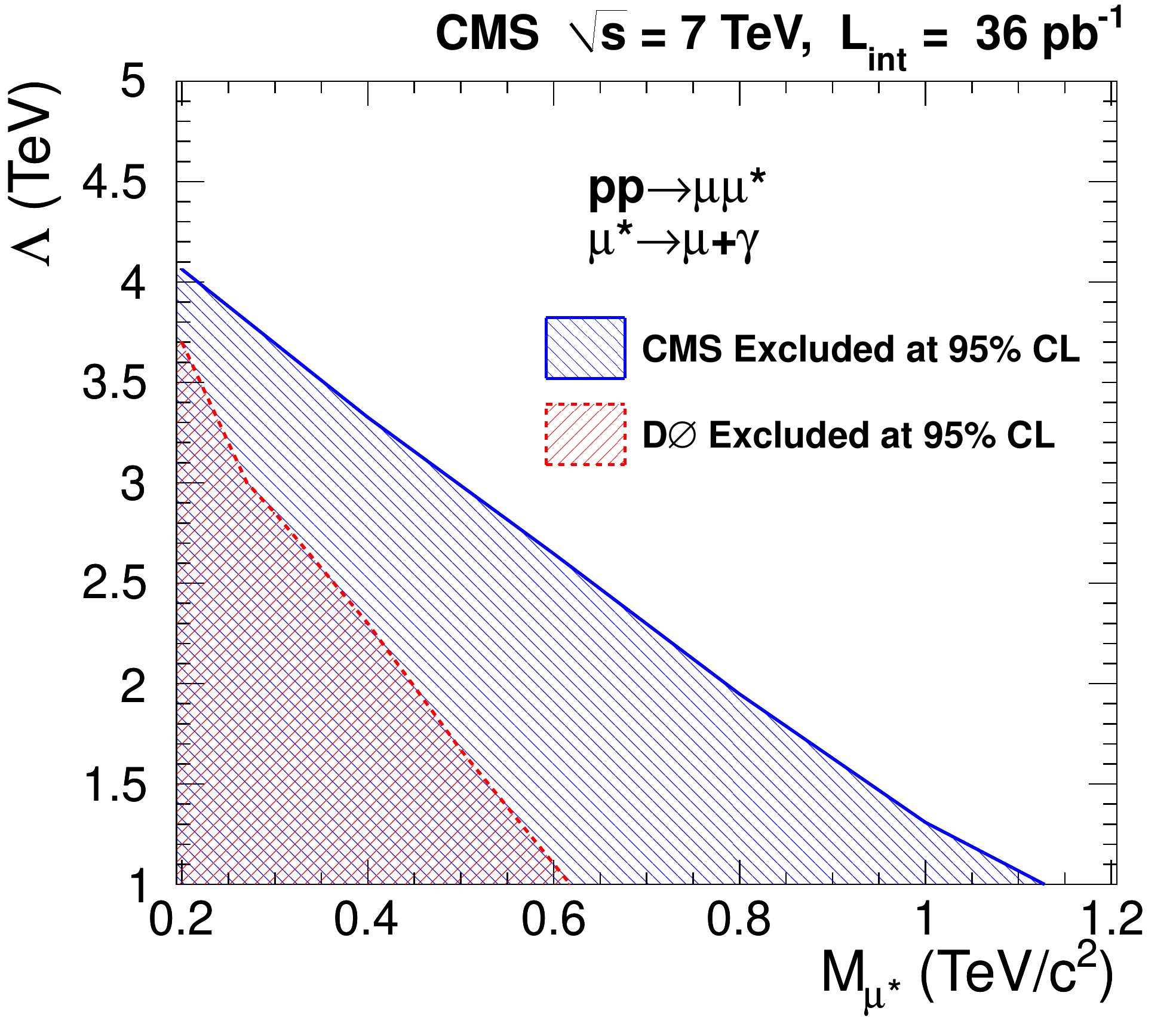}

    \includegraphics[angle=0,width=0.49\linewidth]{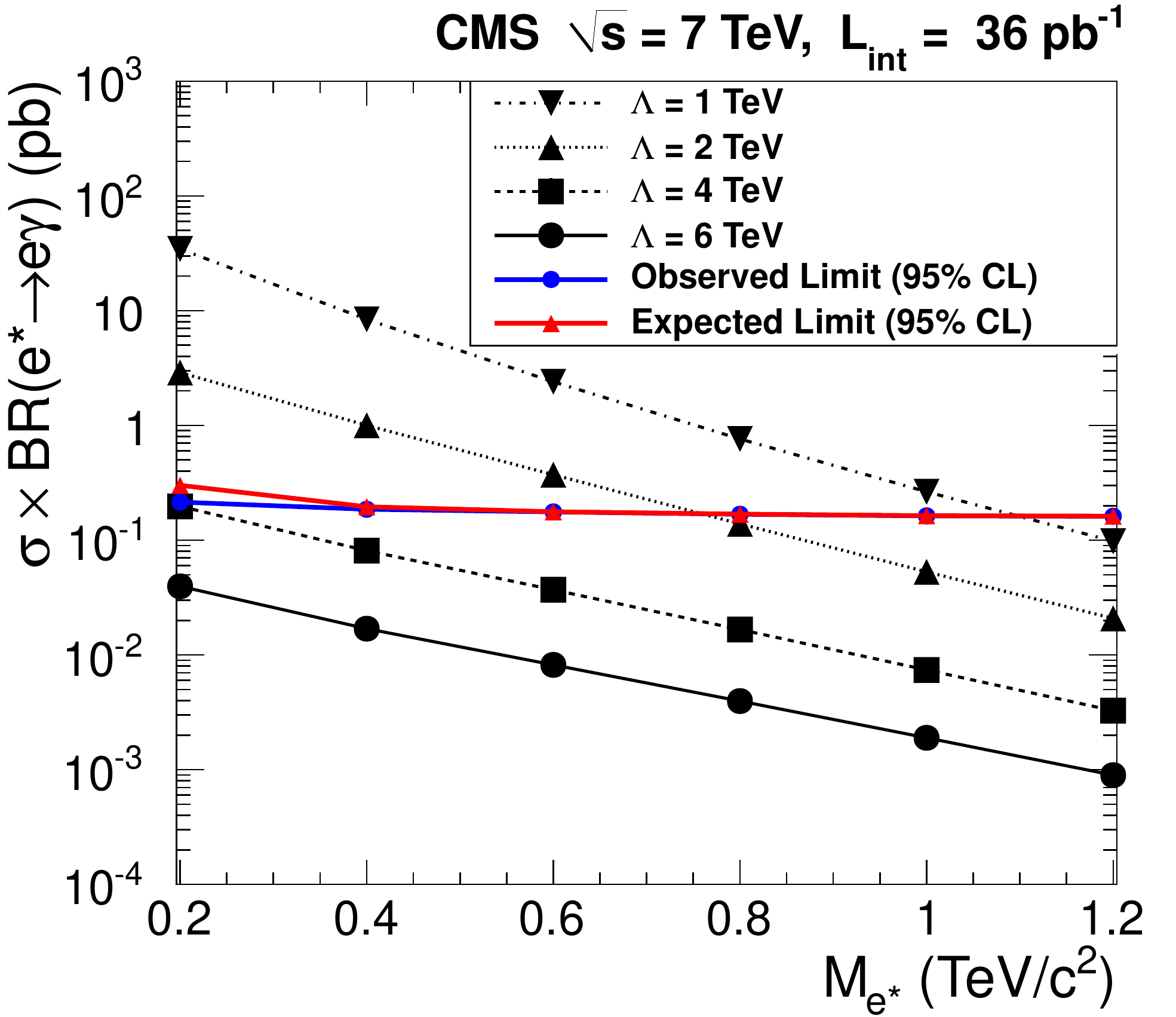}
    \includegraphics[width=0.49\linewidth]{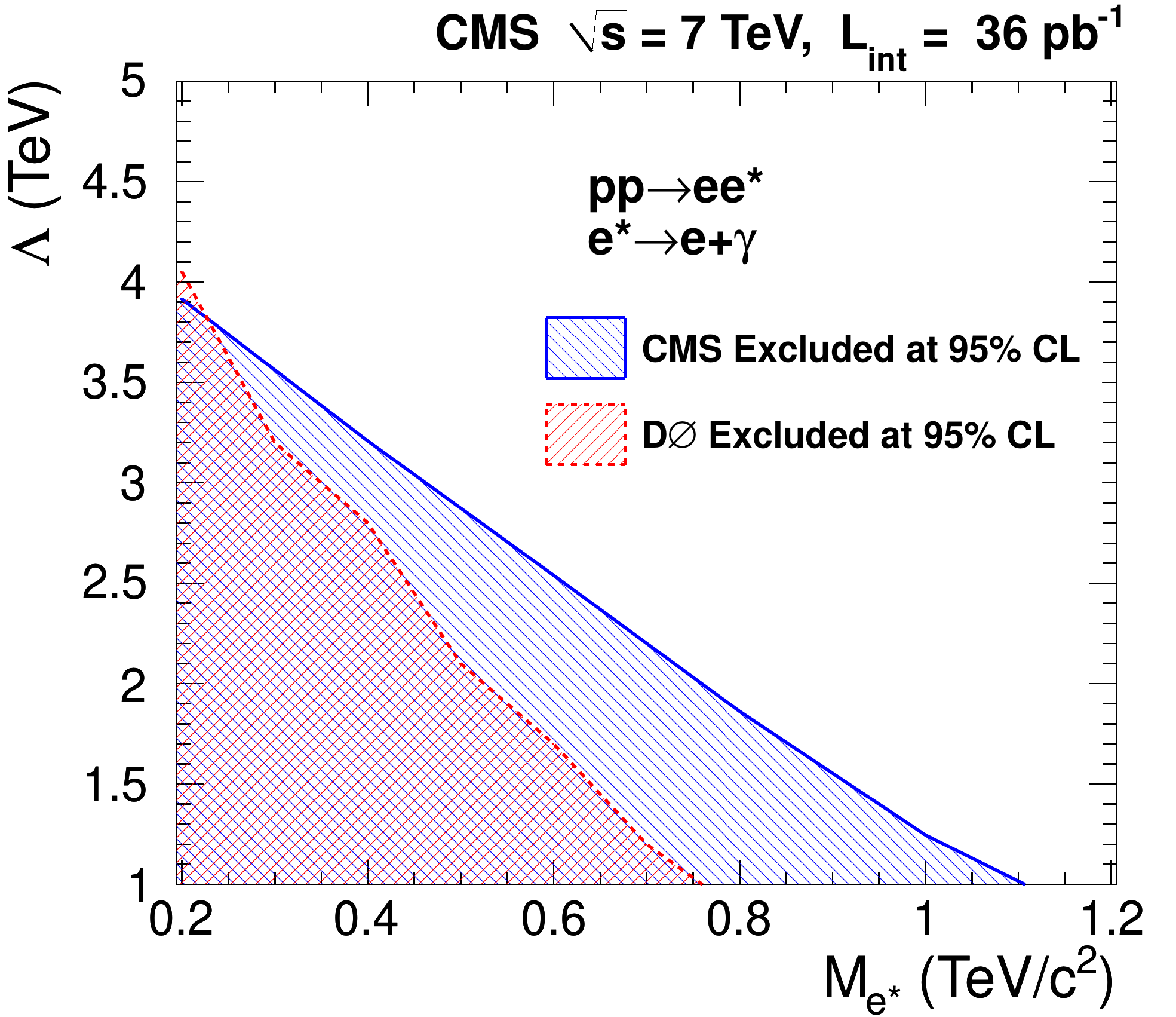}
\caption{
Left: Observed  and expected  limits on the excited muon (top) and electron (bottom)
 production cross section
times branching fraction at 95\% CL, as functions of the excited lepton mass.
The predictions for different $\Lambda$ values are also shown.
Right: The region in the $(\Lambda , M_{\ell^*})$ plane excluded
 at the 95\%~CL by the present search, both for muons (top) and electrons (bottom).
The previous most stringent limits from the D0 Collaboration~\cite{d0,D0estar} are also displayed.
}
\label{fig:limits}
\end{center}
\end{figure*}

Considering the production of excited leptons via a four-fermion
contact interaction as an alternative hypothesis to the SM,
upper limits on the $\ell^*$ production cross section times branching
fraction of the $\ell^* \rightarrow \ell \gamma$ decay are set using
a Bayesian method with a flat prior~\cite{bayes}.
A log-normal prior is used for the integration over the nuisance
parameters. The systematic uncertainties discussed in the previous section are
taken into account in the statistical analysis.
The corresponding expected limit is computed as the weighted average of limits over all
possible numbers of observed events, where the weight is the Poisson probability to
observe a given number of events in data assuming  background-only hypothesis.
The systematic uncertainties discussed in the previous section are taken into account
 in the statistical analysis of the data.
 Cross sections higher than 0.16~pb to 0.21~pb for $\mathrm{e}^*$ production and higher
than 0.14~pb to 0.19~pb
for $\mu^*$ production are excluded at the 95\%
confidence level (CL) for excited lepton masses ranging from 200~\GeVcc to 1500~\GeVcc,
as shown in Fig.~\ref{fig:limits} (left) and given in Table \ref{tab:muresults}.
At  a contact interaction scale of $\Lambda = M_{\ell^*}$,
excited lepton masses are excluded below 1070~\GeVcc
for electrons and 1090~\GeVcc
for muons. If a higher  contact interaction scale  $\Lambda = 2~\TeV$ is considered,
excited lepton masses are excluded below 760~\GeVcc for electrons
and 780~\GeVcc for muons.
Figure~\ref{fig:limits} (right) displays the exclusion regions in
the $(\Lambda, M_{\ell^*})$ plane obtained from these limits,
showing an improvement with respect to the previous most stringent limits established at
hadron colliders~\cite{d0,D0estar}.

\section{Summary}

We have searched for evidence of lepton compositeness  in proton-proton
collisions by looking for production of
excited leptons followed by decay to a lepton and a photon at
$\sqrt{s} = 7 \TeV$. The data sample corresponds to
an integrated luminosity of $36~\mathrm{pb}^{-1}$ collected with the CMS detector.
No excess of events in the $\ell^+\ell^-\gamma$ final state was found
above the SM expectation in the electron or muon channel.
We report the first upper limits
on $\ell^*$ production at this collision energy and exclude a new region of the  $\Lambda - M_{\ell^*}$
parameter space.  At  a contact interaction scale of $\Lambda = 2~\TeV$, excited lepton masses are
excluded at the 95\% CL below 760~\GeVcc for electrons
and 780~\GeVcc for muons.
Assuming $\Lambda = M_{\ell^*} $ instead, excited lepton masses
are excluded below 1070~\GeVcc for electrons
and 1090~\GeVcc for muons, representing the most stringent limits to date.

\section*{Acknowledgments \label{s:ack}}
We wish to congratulate our colleagues in the CERN accelerator departments for
the excellent performance of the LHC machine. We thank the technical and
administrative staff at CERN and other CMS institutes, and acknowledge support
from: FMSR (Austria); FNRS and FWO (Belgium); CNPq, CAPES, FAPERJ, and FAPESP
(Brazil); MES (Bulgaria); CERN; CAS, MoST, and NSFC (China); COLCIENCIAS
(Colombia); MSES (Croatia); RPF (Cyprus); Academy of Sciences and NICPB
(Estonia); Academy of Finland, ME, and HIP (Finland); CEA and CNRS/IN2P3
(France); BMBF, DFG, and HGF (Germany); GSRT (Greece); OTKA and NKTH (Hungary);
CSIR, DAE, and DST (India); IPM (Iran); SFI (Ireland); INFN (Italy); NRF (Korea); LAS
(Lithuania); CINVESTAV, CONACYT, SEP, and UASLP-FAI (Mexico); PAEC (Pakistan);
SCSR (Poland); FCT (Portugal); JINR (Armenia, Belarus, Georgia, Ukraine,
    Uzbekistan); MST and MAE (Russia); MSTDS (Serbia); MICINN and CPAN (Spain);
Swiss Funding Agencies (Switzerland); NSC (Taipei); TUBITAK and TAEK (Turkey);
STFC (United Kingdom); DOE and NSF (USA).

\bibliography{auto_generated}   

\cleardoublepage \appendix\section{The CMS Collaboration \label{app:collab}}\begin{sloppypar}\hyphenpenalty=5000\widowpenalty=500\clubpenalty=5000\textbf{Yerevan Physics Institute,  Yerevan,  Armenia}\\*[0pt]
S.~Chatrchyan, V.~Khachatryan, A.M.~Sirunyan, A.~Tumasyan
\vskip\cmsinstskip
\textbf{Institut f\"{u}r Hochenergiephysik der OeAW,  Wien,  Austria}\\*[0pt]
W.~Adam, T.~Bergauer, M.~Dragicevic, J.~Er\"{o}, C.~Fabjan, M.~Friedl, R.~Fr\"{u}hwirth, V.M.~Ghete, J.~Hammer\cmsAuthorMark{1}, S.~H\"{a}nsel, M.~Hoch, N.~H\"{o}rmann, J.~Hrubec, M.~Jeitler, W.~Kiesenhofer, M.~Krammer, D.~Liko, I.~Mikulec, M.~Pernicka, H.~Rohringer, R.~Sch\"{o}fbeck, J.~Strauss, A.~Taurok, F.~Teischinger, P.~Wagner, W.~Waltenberger, G.~Walzel, E.~Widl, C.-E.~Wulz
% ...
\vskip\cmsinstskip
% ...
1:~~Also at CERN, European Organization for Nuclear Research, Geneva, Switzerland\\
% ...

\end{sloppypar}
\end{document}